\documentclass[final,english]{bullsrsl}

\usepackage[latin1]{inputenc}
\usepackage[T1]{fontenc}
\usepackage{natbib}
\usepackage{graphicx}

\begin{document}

\title{Characterisation of the ILMT 4K $\times$ 4K TDI CCD imager}

\author[affil={1}, corresponding]{Gokul Singh}{Mehra}
\author[affil={2,1}]{Bhavya}{Ailawadhi}
\author[affil={3,4}]{Talat}{Akhunov}
\author[affil={1,5}]{Monalisa}{Dubey}
\author[affil={1,5}]{Naveen}{Dukiya}
\author[affil={6}]{Sara}{Filali}
\author[affil={7,8}]{Paul}{Hickson}
\author[affil={1}]{Priyanshi}{Kumari}
\author[affil={1}]{Kuntal}{Misra}
\author[affil={9}]{Vibhore}{Negi}
\author[affil={1,10}]{Kumar}{Pranshu}
\author[affil={6,1}]{Jean}{Surdej}
\author[affil={1,10}]{Sarvesh Kumar}{Yadav}

\affiliation[1]{Aryabhatta Research Institute of Observational Sciences (ARIES), Manora Peak, Nainital-263002, India}
\affiliation[2]{Physical Research Laboratory, Navrangpura, Ahmedabad, Gujarat-380009, India}
\affiliation[3]{National University of Uzbekistan, Department of Astronomy and Astrophysics, 100174 Tashkent, Uzbekistan}
\affiliation[4]{Ulugh Beg Astronomical Institute of the Uzbek Academy of Sciences, Astronomicheskaya 33, 100052 Tashkent, Uzbekistan}
\affiliation[5]{Mahatma Jyotiba Phule Rohilkhand University, Pilibhit Bypass Road, Bareilly 243006, Uttar Pradesh, India
}
\affiliation[6]{Institute of Astrophysics and Geophysics, Li\`ege University, All\'ee du 6 Ao\^ut 19c, 4000 Li\`ege, Belgium
}
\affiliation[7]{Department of Physics and Astronomy, The University of British Columbia, 6224 Agricultural Road, Vancouver, BC V6T 1Z1, Canada
}

\affiliation[8]{Outer Space Institute, The University of British Columbia, 325-6224 Agricultural Road, Vancouver, BC V6T 1Z1, Canada
}

\affiliation[9]{Kavli Institute for Astronomy and Astrophysics, Peking University, Beijing 100871, China}

\affiliation[10]{Department of Applied Optics and Photonics, University of Calcutta, Kolkata, 700106, India}

\correspondance{gokulmehra51@gmail.com}

\maketitle

\begin{abstract}
In this work, we determine the gain and read-out noise of the 4K$\times$4K CCD imager used with the 4-m International Liquid Mirror Telescope (ILMT). Due to the zenith-pointing characteristic of the telescope, the CCD operates in the Time Delay Integration (TDI) mode for image acquisition. The CCD frames obtained during rising twilight have non-uniform illumination and can potentially be used to calculate important CCD parameters using two approaches. The first approach directly measures variance in flux along rows with minimum horizontal gradient, despite the strong vertical illumination gradient. The second method divides the CCD frame into regions exposed to similar background illumination and analyses the flux ratio variation between the two regions. Mathematically, we establish a linear relationship between the signal variance and average flux, from which gain is determined, while the read-out noise (RON) is measured independently from the bias frames. Both methods provide a gain value of approximately 3.9 e/ADU, while the RON measured from bias frames is 1.55 ADU ($\simeq 6\, e$). Estimation of these parameters provides a better understanding of the CCD performance and allows for an accurate calibration.

\end{abstract}


\section{Introduction}

The International Liquid Mirror Telescope is a 4-m aperture, zenith-pointing telescope located at the Devasthal Observatory of ARIES at an altitude of approximately 2378~m. ILMT uses liquid mercury as its primary reflecting surface. 
The mirror is formed by pouring mercury into a bowl-shaped container and rotating it about the vertical axis with a period of 8.02~s, forming a stable parabolic reflecting surface. It has a field of view 22-arcminutes and scans the same strip of sky each observing night, and employs the SDSS $g^{\prime}$, $r^{\prime}$, and $i^{\prime}$ filters \citep{2025A&A...694A..80S}.

In this paper, we determine two key CCD (charge-coupled device) parameters, the gain (G) and RON, using two complementary methods. CCDs convert incoming photons into measurable electrical signals and have revolutionized astronomical observations. Their performance is characterized by several parameters, among which gain and RON are fundamental for accurate scientific measurements \citep{2001Janesick,Howell2006}. The gain defines the conversion factor between the number of photoelectrons collected in a pixel and the recorded digital counts, while the RON quantifies the uncertainty introduced during the readout process \citep{2001Janesick}.

The paper is organised as follows: Section~\ref{Methodology} presents the theoretical background  of the noise model, data used and two independent methods employed for the computation of CCD parameters. The results and further discussions are presented in Sections~\ref{Results} and~\ref{Discussions}, respectively, and the conclusions are summarized in Section~\ref{Conclusion}.

\section{Methodology}
\label{Methodology}

\subsection{Theoretical Background}
During an exposure, a photon hitting a detector produces a photoelectron, and at the end of the exposure, the charges (photoelectrons) are read out and converted to a digital signal. This digital value is termed ADU (analog-to-digital units). The relation between the flux measured in electrons F(e) and flux in ADU is given by the simple linear relation \citep{Howell2006}.

\begin{equation}
\label{eq:1}
F(e) = F(\mathrm{ADU}) \times G(\mathrm{e/ADU}),
\end{equation}
where G is the Gain factor. The total noise (variance) in a single pixel arises from multiple sources: the Poisson noise from the signal itself $F(e)$, and the Gaussian RON $\sigma_{\mathrm{RON}}(e)$. The combined variance is:
\begin{equation}
\label{eq:2}
\sigma^2(e) = F(e) + \sigma_{\mathrm{RON}}^2(e).
\end{equation}
Converting this relationship to ADU units with the help of Eq.~\ref{eq:1}, we obtain:
\begin{equation}
\label{eq:3}
\sigma^2(\mathrm{ADU}) = 
\frac{F(ADU)}{G(e/ADU)}+\sigma^2_{RON}(ADU).
\end{equation}

This equation predicts a linear relationship between the variance and the mean flux F(ADU), where the reciprocal of the slope gives the Gain and the square root of the intercept gives the RON.

\subsection{Data used}
\begin{figure}[h]
    \centering
    \includegraphics[width=0.50\linewidth]{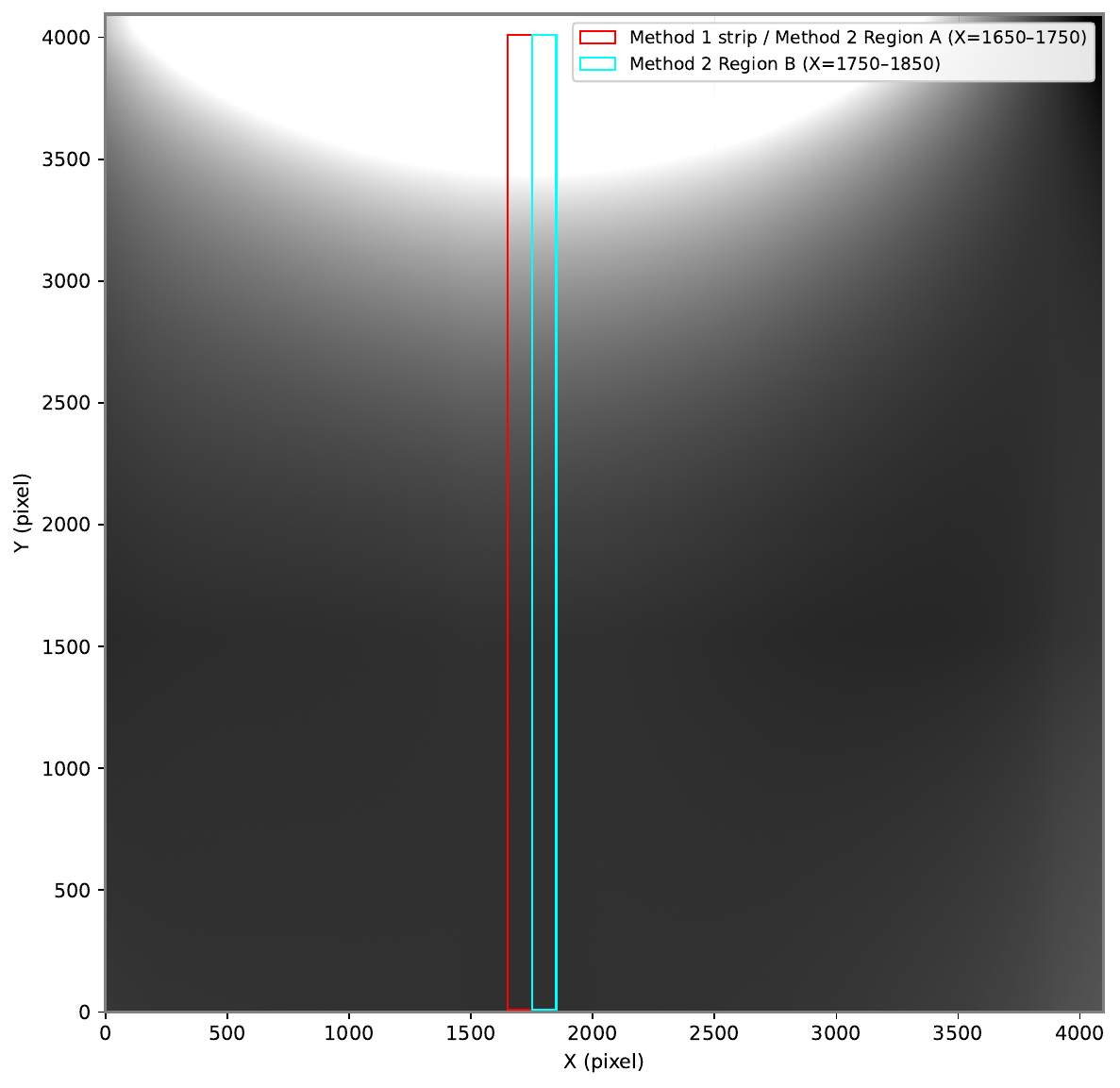}
    
    \begin{minipage}{12cm}
    \caption{CCD frame obtained during twilight.}
    \label{fig:CCD_frame} 
    \end{minipage}

\end{figure}

The data were obtained during rising (dawn) twilight on 2024 October 29, with the CCD operating in TDI mode at a temperature of $-98.5\,^\circ\mathrm{C}$ and a frame size of $4096 \times 4096$ pixels. A single twilight frame (r$'$ filter) was selected for analysis and corrected for dark current and flat-field response using 1-D master calibration frames \citep{2025A&A...694A..80S}, each built from five independent sets of observations. Since the dark frames share the science and flat-field exposure time, they already include the bias offset, removing the need for separate bias subtraction. The corrected frame (Figure~\ref{fig:CCD_frame}) shows vertical and horizontal flux gradients from increasing sky brightness and vignetting, respectively; two complementary methods estimate the gain in the presence of these gradients (Sections~2.3 and~2.4).

\subsection{Method 1: Direct Variance Method}
The first method attempts to minimize the effect of the horizontal gradient. For this the following steps were followed:
\begin{itemize}
    \item We identified a region of the CCD where the horizontal gradient is minimal around X=1700. We selected a narrow region 100 pixels wide and then to capture the vertical
    gradient in this region we extended from row 10 to 4000 (along Y-axis).
    
    \item For each row within this strip, we calculate the average flux $F(\mathrm{ADU})$ and the variance $\sigma^2(\mathrm{ADU})$ and plot the two quantities; according to Eq.~\ref{eq:3}, this should produce a linear relationship, from which we can extracted the slope ($1/G$) and intercept.
\end{itemize}
Since the lowest flux recorded in this exposure was not sufficiently small, we could not determine the RON from these observations.

\subsection{Method 2: Two-Region Ratio Method}

In the second method, we measured the signal variance in two regions of nearly equal mean flux. Taking the ratio of the fluxes in these regions largely cancels out the illumination gradient, and the remaining variation in the ratio is due to photon noise and RON of the CCD electronics. Assuming the flux and its variance are approximately equal in the two regions, $F_1 \approx F_2 \approx F$ and $\sigma(F_1) \approx \sigma(F_2) \approx \sigma(F)$, we derive the following relation through error propagation:

\begin{equation}
    \sigma(F) = \frac{F}{\sqrt{2}} \, \sigma(F_1/F_2),
    \label{eq:4}
\end{equation}

with the help of Eq.~\ref{eq:4}, we can rewrite Eq.~\ref{eq:3} as:
\begin{equation}
\label{eq:5}
    \frac{F^2}{2} \, \sigma^2(F_1/F_2) = 
    \frac{F \, (\mathrm{ADU})}{G \,\mathrm{(e/ADU)}} + \sigma^2_{\mathrm{RON}} \, (\mathrm{ADU}).
\end{equation}

Eq.~\ref{eq:5} has the same linear form as Eq.~\ref{eq:3}, which allows us to plot the quantity $\frac{F^2}{2}$ $\times$ $\sigma^2(F_1/F_2)$ against F in order to evaluate the Gain from the slope $1/G(e/ADU)$.

\section{Results}
\label{Results}

\begin{figure}[h]
    \centering
    \includegraphics[width=0.48\textwidth]{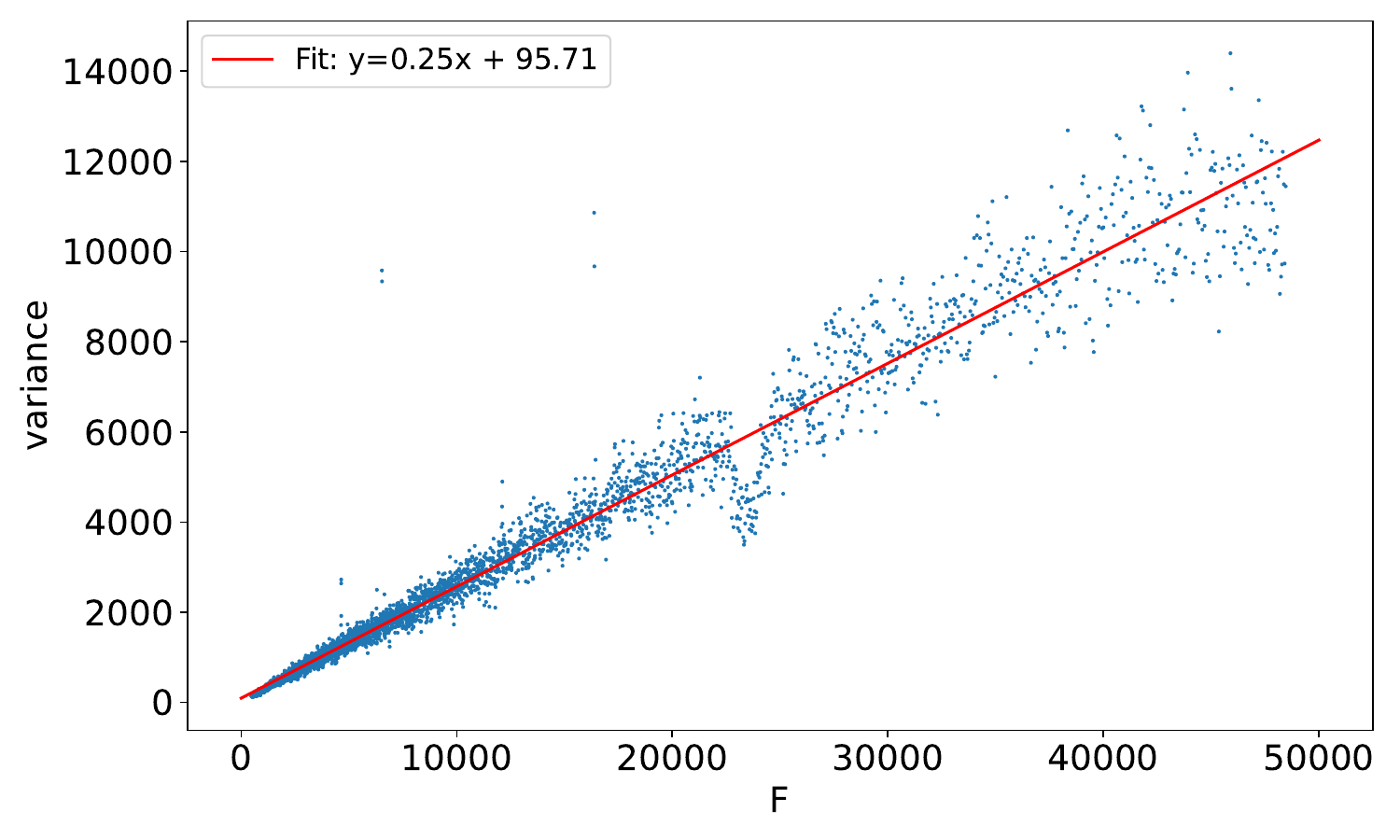}\hfill
    \includegraphics[width=0.48\textwidth]{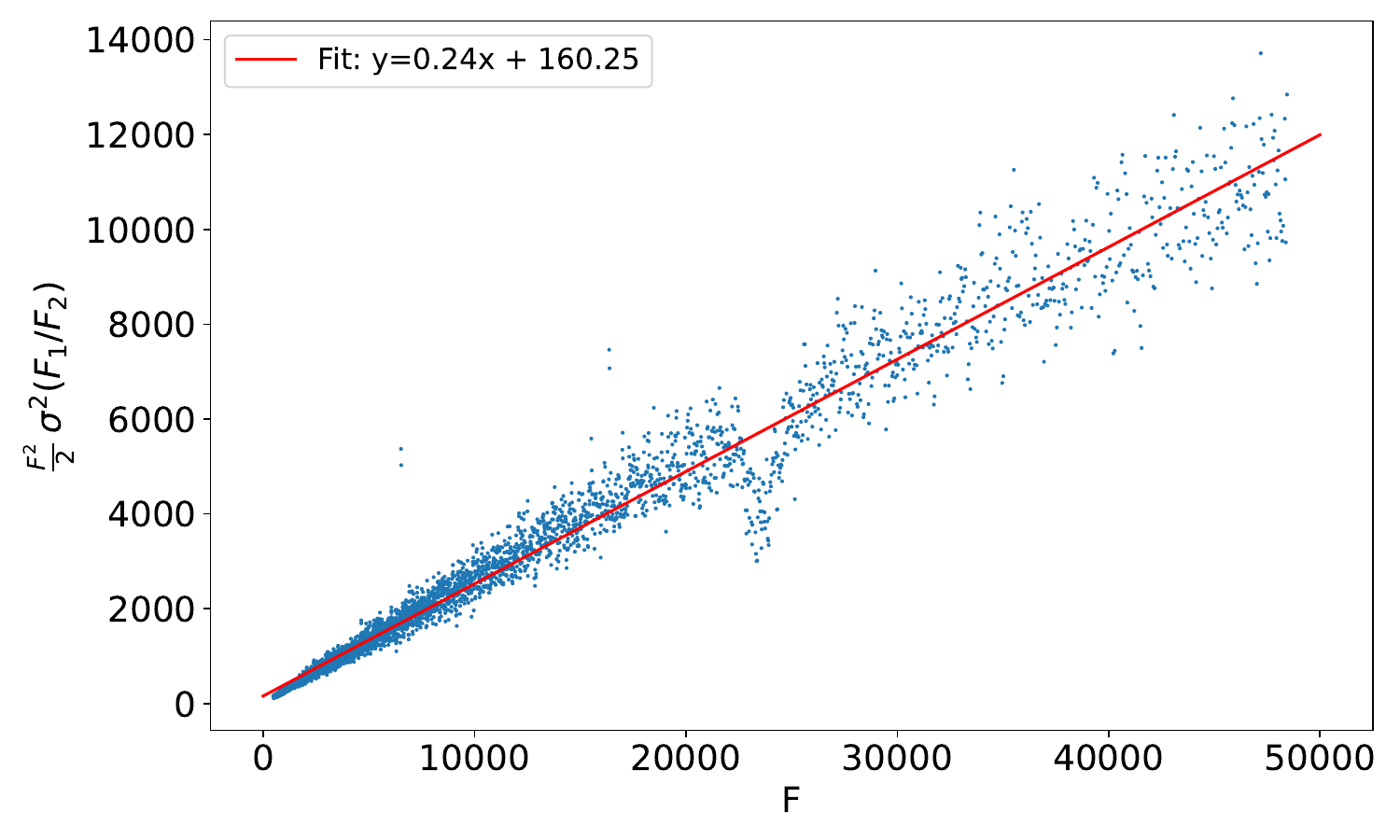}
    
    \begin{minipage}{12cm}
    \caption{Variance-flux plots from the twilight CCD frames for the two methods. \textbf{Left}:$\sigma^2(\mathrm{ADU})$ vs.\ $F$ for method~1; the fitted slope of $0.25$ gives $G \simeq 4\,e/\mathrm{ADU}$ (Eq.~\ref{eq:3}). \textbf{Right}:$\frac{F^2}{2}\,\sigma^2(F_1/F_2)$ vs.\ $F$ for method~2; the fitted slope of $0.24$ gives $G \simeq 4.2\,e/\mathrm{ADU}$ (Eq.~\ref{eq:5}).}
    \label{fig:2}
    
    \end{minipage}
\end{figure}

Following Method 1 and Method 2, we plot the measured variance and mean flux according to Eq.~\ref{eq:3} and Eq.~\ref{eq:5}, and fit a straight line to the data to get the slope from the linear relationship. The plots for both methods are shown in Figure~\ref{fig:2}. Since the twilight frames do not record sufficiently low flux, the fitted intercepts from Eqs.~\ref{eq:3} and~\ref{eq:5} cannot be reliably used to estimate the RON, as the linear fit is extrapolated to zero flux from data that does not sample the low-flux regime. These intercepts, $\sqrt{95.71} \approx 9.8~\mathrm{ADU}$ and $\sqrt{160.25} \approx 12.7~\mathrm{ADU}$, are both substantially higher than the RON obtained independently from bias frames.

The RON was therefore calculated separately from a set of 10 bias frames. For these near-zero exposure frames, after subtracting the average bias value, the pixel signal is centered near zero and follows a Gaussian distribution, whose width gives the RON in ADU, $\sigma_{\mathrm{RON}}(\mathrm{ADU})  \approx 1.55~\mathrm{ADU}$. Using the gain, this corresponds to $\sigma_{\mathrm{RON}}(e^-) \approx 6~e^-$.

\begin{figure}[h]

    \centering

    \includegraphics[width=0.48\textwidth]{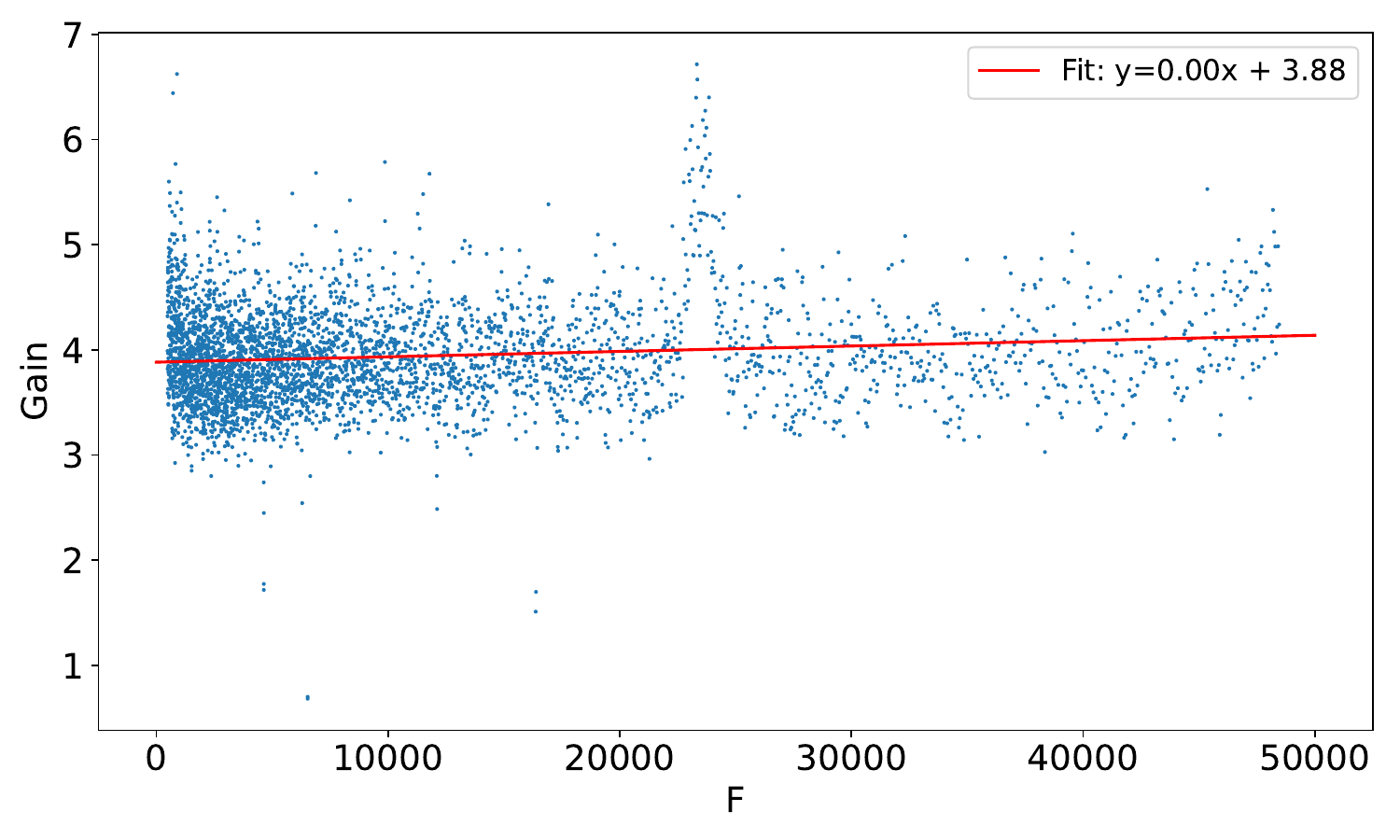}\hfill
    \includegraphics[width=0.48\textwidth]{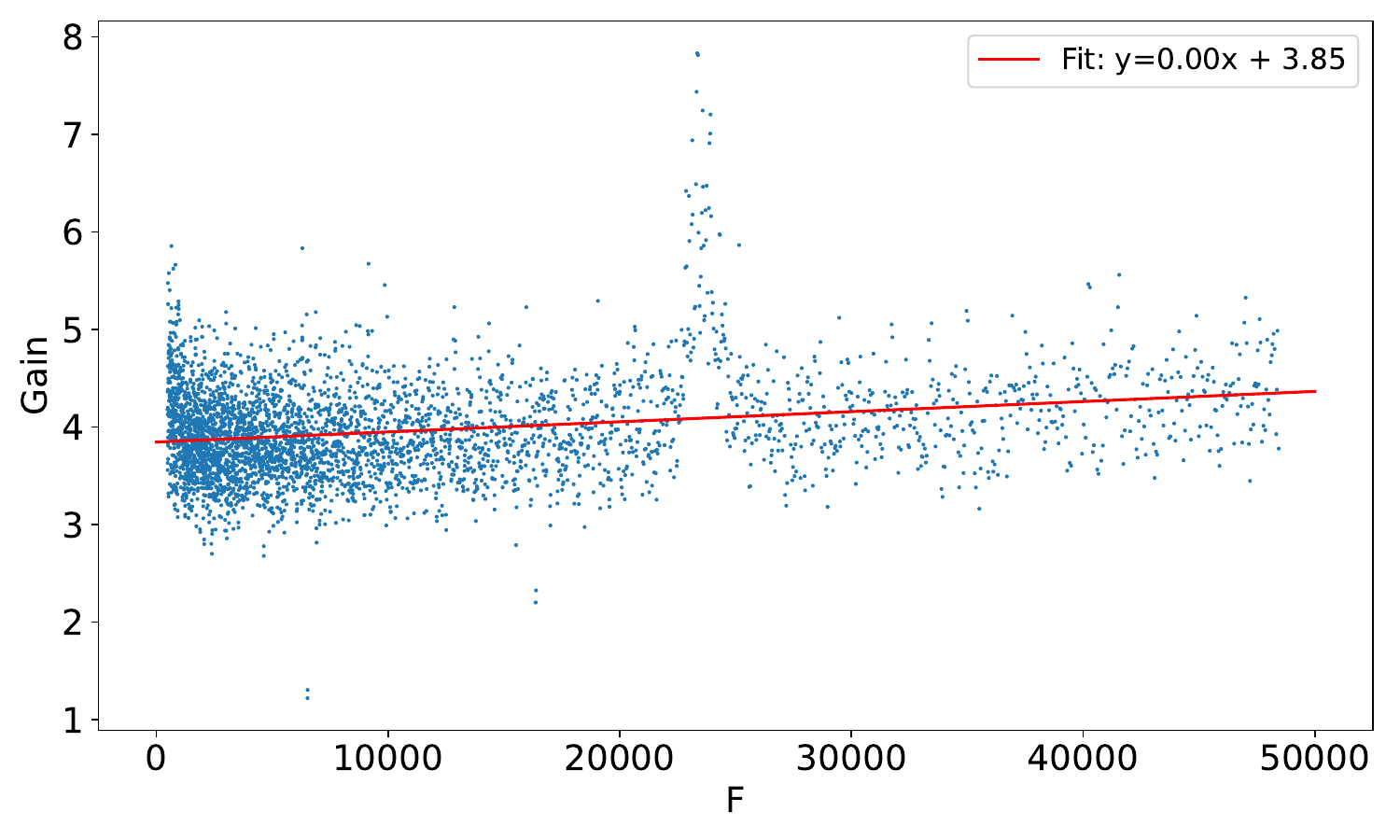}
    
    \begin{minipage}{12cm}
    \caption{Gain-flux plots from the twilight CCD frame for the two methods. \textbf{Left}:Gain vs.\ $F$ for method~1, showing a constant trend; the linear fit gives $G \simeq 3.9$. \textbf{Right}:Gain vs.\ $F$ ($F = \frac{F_1+F_2}{2}$) for method~2, again follows a constant trend; the linear fit gives $G \simeq 3.8$.}

    \label{fig:3}
\end{minipage}

\end{figure}

Using the average flux, per-row variance, and $\sigma_{RON}(\mathrm{ADU}) \approx 1.55~\mathrm{ADU}$, we calculate the gain for each row via Eqs.~\ref{eq:3} and~\ref{eq:5}, and plot it against average flux to check consistency across brightness levels. Both panels in Figure~\ref{fig:3} show a similar trend, yielding a gain of $\approx 3.98$.

\section{Discussions}
\label{Discussions}

Figures~\ref{fig:2} and~\ref{fig:3} show the measured variance-signal and gain-signal relationships. The data follow an essentially linear trend over the fitted range (5000-30000 ADU), yielding the gain estimates reported in Section 3. A noticeable increase in scatter appears at intermediate fluxes but does not significantly affect the derived gain or the overall linear trend.

\section{Conclusion}
\label{Conclusion}

We have characterized the ILMT $4K\times4K$ CCD using two independent flux-variance methods. Both methods yield a gain of $\approx3.9~(e/{\mathrm{ADU}})$. The RON calculated from the bias frames is $1.55 \ {\mathrm{ADU}}$. The agreement between the results obtained from two independent methods supports the methodology adopted.

\begin{acknowledgments}
The 4-m International Liquid Mirror Telescope (ILMT) project results from a collaboration between the Institute of Astrophysics and Geophysics (University of Li\`{e}ge, Belgium), the Universities of British Columbia, Laval, Montreal, Toronto, Victoria, and York University, and Aryabhatta Research Institute of Observational Sciences (ARIES, India). The team acknowledges the contributions of ARIES's technical and scientific staff. JS wishes to thank Service Public Wallonie, F.R.S.-FNRS (Belgium), and the University of Li\`{e}ge, Belgium, for funding the construction of the ILMT. PH acknowledges financial support from the Natural Sciences and Engineering Research Council of Canada, RGPIN-2019-04369. PH, SF, and JS thank ARIES for hospitality during their visits to Devasthal. We sincerely acknowledge Anna Pospiezalska for her significant contributions to ILMT functioning. MD acknowledges the Innovation in Science Pursuit for Inspired Research (INSPIRE) fellowship award (DST/INSPIRE Fellowship/2020/IF200251) for this work. KP acknowledges the support from the Erasmus+ Programme of the European Union for a research visit to the Institute of Astrophysics and Geophysics, University of Li\`ege, Belgium. TA thanks the Ministry of Higher Education, Science, and Innovations of Uzbekistan (grant FZ-20200929344). JS and KM acknowledge the assistance received from the ANRF SERB-726 VAJRA Faculty Scheme, India. KM acknowledges the support from the BRICS grant DST/ICD/BRICS/Call-5/CoNMuTraMO/2023 (G) funded by the DST, India.
\end{acknowledgments}

\begin{furtherinformation}
\begin{authorcontributions}
All authors contributed significantly to the work presented in this paper.
\end{authorcontributions}

\begin{conflictsofinterest}
The authors declare no conflict of interest.
\end{conflictsofinterest}
\end{furtherinformation}
\bibliographystyle{bullsrsl-en.bst}

\bibliography{BINA_gokul}
\end{document}